\documentclass{llncs}
\usepackage{makeidx}  
\usepackage{graphicx}
\usepackage{balance}  
\usepackage{url}

\usepackage[normalem]{ulem}
\usepackage{algorithm}
\usepackage{verbatim}
\usepackage{color}
\usepackage{cite}
\def\poss{\mathop{\mit Poss}}
\def\cert{\mathop{\mit Cert}}
 
\begin{document}
\frontmatter          
\pagestyle{headings}  

\title{Conditional Tables in practice}
\titlerunning{PossDB}  
%
\author{G\"{o}sta Grahne \and Adrian Onet
\and Nihat Tartal }
%
%
%
\institute{Concordia University, Montreal, QC, H3G 1M8, CANADA,\\
\email{\{grahne,a\_onet,m\_tartal\}@cs.concordia.ca}}

\maketitle              

\begin{abstract}
Due to the ever increasing importance of
the internet, interoperability of
heterogeneous data sources
is as well of ever increasing
importance. Interoperability can be achieved
e.g.\ through data integration and data exchange.
Common to both approaches is the need for the
DBMS to be able to store and query
{\em incomplete databases}. In this report
we present PossDB, a DBMS
capable of storing and querying incomplete databases.
The system is wrapper over PostgreSQL, and the query language
is an extension of a subset of standard SQL.
Our experimental results show that our
system scales well, actually better than
comparable systems.
\end{abstract}
\section{Introduction}
Management of uncertain and imprecise data has long
been recognized as an important direction
of research in data bases.
With the tremendous growth of
information stored and shared over the Internet,
and the introduction of new technologies
able to capture and transmit information,
it has become increasingly important for
Data Base Management Systems
to be able to
handle uncertain and probabilistic data.
As a consequence, there has lately been
significant efforts by the database
research community to develop
new systems able to deal with
uncertainty,
either by annotating
values with probabilistic measures
(e.g.\
MystiQ \cite{DBLP:conf/sigmod/BoulosDMMRS05},
Orion \cite{DBLP:conf/comad/SinghMMPHS08},
BayesStore \cite{DBLP:journals/pvldb/WangMGH08},
PrDB \cite{DBLP:journals/vldb/SenDG09},
MayBMS-2 \cite{DBLP:conf/icde/AntovaKO07a})
or
defining new structures capable of capturing
missing information (e.g.\ Trio \cite{ilprints658}
and
MayBMS-1 \cite{Antova:2009:WBE:1644245.1644253}).

Uncertainty management is an
important topic also in data exchange and information integration.
In these scenarios the data stored in one database
has to be restructured to fit the schema of a different
database.
The restructuring forces the introduction of ``null'' values
in the translated data, since the second schema can
contain columns not present in the first.
In the currently commercially
available relational
DBMS's the missing or unknown information
is stored with placeholder
values denoted {\tt null}.
It is well known that
this representation
has  drawbacks
when it comes to
query answering,
and that a logically coherent treatment
of the {\tt null} is still lacking
from most DBMS's.

To illustrate the above mentioned drawbacks,
consider a merger of companies
``Acme"  and ``Ajax."
Both companies keep an employee database.
Let
$Emp1(\mbox{\em Name,Mstat,Dept})$ and
$Emp2(\mbox{\em Name,Gender, Mstat})$,
where {\em Mstat} stands for marital status,
be the schemas
used by Acme and Ajax, respectively.
The merged company decides to use the schema
$Emp(\mbox{\em Name,Gender,Mstat,Dept})$, and
it is known that all
the employees from Ajax will work under the same
department, which will either be 'IT' or 'PR'.
Consider now the initial data from both companies:

\medskip
\begin{minipage}[b]{0.48\linewidth}
\centering
\begin{tabular}{lll}
\multicolumn{3}{c}{Emp1}    \\ \hline
Name     & Mstat      & Dept \\ \hline\hline 
Alice    & married    & IT   \\
Bob      & married    & HR   \\
Cecilia  & married    & HR
\end{tabular}
\end{minipage}
\begin{minipage}[b]{0.45\linewidth}
\centering
\begin{tabular}{lll}
\multicolumn{3}{c}{Emp2} \\ \hline
Name  & Gender & Mstat\\ \hline\hline 
David  & M & married \\
Ella   & F & single\\ \\
\end{tabular}
\end{minipage}
\medskip

The merged company database instance would be
represented as the following
database in a standard relational DBMS:

\begin{center}
\begin{tabular}{llll}
\multicolumn{4}{c}{Emp}                      \\ \hline
Name    & Gender     & Mstat & Dept        \\ \hline\hline 
Alice   & {\tt null} & married & IT          \\
Bob     & {\tt null} & married & HR          \\
Cecilia & {\tt null} & married & HR          \\
David   & M          & married & {\tt null}  \\
Ella    & F          & single  & {\tt null}
\end{tabular}
\end{center}

With this incomplete database consider now the following
two simple queries:

\medskip
\noindent
$Q_1$: {\small{\tt SELECT Name FROM Emp  WHERE}}

{\small{\tt(Gender = 'M' AND Mstat = 'married') OR Gender = 'F'}}

\medskip
\noindent
$Q_2$: {\small{\tt SELECT E.Name, F.Name FROM Emp E, Emp F}}

\noindent
\hspace{2cm}{\small{\tt  WHERE E.Dept=F.Dept AND E.Name != F.Name}}
\medskip

Clearly one would expect that the first query to return
all employee names and the second query to return the set of tuples
$\{(\mbox{\em Bob, Cecilia}),(\mbox{\em David, Ella})\}$.
Unfortunately by the default way null values
are treated in standard systems the tuples returned by the
first query would return the set
$\{(\mbox{\em David, Ella})\}$
and by the second query would return the set
$\{(\mbox{\em Bob, Cecilia})\}$.

In this report we introduce a new database management system
called PossDB
(Possibility Data Base) able to
fully support incomplete information.
The purpose of the PossDB system
is to demonstrate that scalable processing
of semantically meaningful null values
is indeed possible, and can be built on
top of a standard DBMS (PostgreSQL, in our case).

Irrespectively of how an incomplete database
instance ${\cal I}$
is represented,
conceptually
it is a (finite or infinite)
{\em set} of possible complete database instances $I$
(i.e.\ databases without null values),
denoted $\poss({\cal I})$.
Each $I\in\poss({\cal I})$ is called a
{\em possible world} of ${\cal I}$.
A query $Q$ over a complete instance $I$ gives a
complete instance $Q(I)$ as answer.
For incomplete databases there are
three semantics for
query answers:

\begin{enumerate}
\item {\em The exact answer}.
The answer is (conceptually) a set of complete
instances, each obtained by querying a
possible world of ${\cal I}$, i.e.\
$\{Q(I) : I\in\poss({\cal I})\}$.
The answer should be represented in the same way
as the input database, e.g.\ as a relation with meaningful nulls. \\

\item {\em The certain answer}.
This answer is a complete database containing
only the (complete) tuples that appear in
in the query answer in {\em all} possible worlds.
In other words,
$\cert(Q({\cal I})) = \bigcap_{I\in\poss({\cal I})}Q(I)$.\\

\item {\em The possible answer}.
$\poss(Q({\cal I})) = \bigcup_{I\in\poss({\cal I})}Q(I)$.
\end{enumerate}

The PossDB system is based on conditional tables
(c-tables)
\cite{lipski:incomplete}
which generalize relations in three ways.
First, in the entries in the columns, variables,
representing unknown values, are allowed in addition
to the usual constants.
The same variable may
occur in several entries,
and it represents the {\em same} unknown
value wherever it occurs.
A c-table $T$ represents a
set of complete instances, each obtained by substituting
each variable with a constant,
that is, applying a valuation $v$ to the table,
where $v$ is a mapping from the variables to constants.
Each valuation $v$ then gives rise to a possible world
$v(T)$.
The second generalization is
that each tuple $t$ is associated with a
{\em local condition} $\varphi(t)$,
which is a Boolean formula over equalities between
constants and variables, or variables and variables.
The final generalization introduces a
{\em global condition} $\Phi(T)$,
which has the same form as the local conditions.
In obtaining complete instances from a table $T$,
we consider only those valuations $v$, for which
$v(\Phi(T))$ evaluates to {\em True},
and include in $v(T)$ only tuples~$v(t)$, where
$v(\varphi(t))$ evaluates to {\em True}.

In our previous example the merged incomplete database
would be represented as the following c-table.

\begin{center}
\begin{tabular}{llll|l}
\multicolumn{5}{c}{Emp} \\ \hline
Name     & Gender & Mstat   & Dept    & $\varphi(t)$  \\ \hline\hline 
Alice   & $x_1$   & married & IT      & {\em True}    \\
Bob     & $x_2$   & married & HR      & {\em True}    \\
Cecilia & $x_3$   & married & HR      & {\em True}    \\
David   & M       & married & $x_4$   & {\em True}   \\
Ella    & F       & single  & $x_4$   & {\em True}
\end{tabular}
\end{center}

The global condition $\Phi(\mbox{Emp})$ 
is $(x_i =\mbox{'M'})\vee(x_i =\mbox{'F'})$,
for $i=1,2,3$, and
$(x_4=\mbox{'IT'})\vee(x_4=\mbox{'PR'})$.
Under this interpretation PossDB will return the
expected results for both queries. Note that in this example
the exact, possible, and certain answers are the same.

The c-tables support the full
relational algebra \cite{lipski:incomplete},
and are capable of
returning the possible, the certain and the
exact answers.
A (complete)
tuple $t$ is in the possible answer to a query $Q$,
if $t\in Q(v(T))$ for {\em some} valuation $v$,
and $t$ is in the certain answer
if $t\in Q(v(T))$ for {\em all} valuations $v$.
The exact answer of a query $Q$ on a
c-table $T$ is a
c-table $Q(T)$
such that $v(Q(T))= Q(v(T))$,
for all valuations $v$.

C-tables are the oldest and most fundamental
instance of a {\em semiring-labeled} database
\cite{Green:2007:PS:1265530.1265535}.
By choosing the appropriate semiring,
labeled databases can model a variety
of phenomena in addition to incomplete information.
Examples are probabilistic databases,
various forms of database provenance,
databases with bag semantics, etc.  
It is our view that the experiences obtained
from the PossDB project will also be applicable
to other semiring based databases.

To the best of our knowledge, PossDB is the first
implemented system based on c-tables.
In the
future we plan to
extend our system to support the
Conditional Chase \cite{Grahne:2011:CWC:1966357.1966360},
a functionality which is highly relevant
in data exchange and information integration.

In order to gauge the scalability of our system,
we have run some experiments comparing the
performance of PossDB with MayBMS-1 \cite{Antova:2009:WBE:1644245.1644253}.
The MayBMS-1 system uses a representation
mechanism called {\em World Set Decompositions},
which is fundamentally different from
c-tables. For details we refer to \cite{Antova:2009:WBE:1644245.1644253}.
Similarly to PossDB, the MayBMS-1 system is build on top of
PostgreSQL, an open source relational database
management system. In the case where there is no
incomplete information, both PossDB and MayBMS-1 work
exactly like classical DBMS's.
However,
at this point we have restricted,
similarly to MayBMS-1, our data
to be encoded as positive integers.
In the future we will
extend the allowed data types to
include all the base data types.

In this current stage PossDB allows the following
operations:

$(1):$ Creation of c-tables,

$(2):$ Querying c-tables,

$(3):$ Inserting into c-tables,

$(4):$ Materializing c-tables representing
       the exact answers to queries, and
      
$(5):$ Testing for tuple possibility and certainty in c-tables.\\

All these operations are expressed using an
extension of the ANSI SQL language
called C-SQL (Conditional SQL).


\section{Features of PossDB}
The PossDB system has system specific
operations and functions
related to c-tables.
To illustrate
these operations,
let us continue with the
example from the
previous section.
The global condition in our example is
\begin{small}
$\Phi(Emp)=_{\sf def}
\{(x_i=\mbox{'M'} \vee x_i=\mbox{'F'}) : i=1,2,3\}
\cup
\{x_4=\mbox{'IT'}\vee x_4=\mbox{'PR'}\}$.
\end{small}
This set corresponds to a CNF formula,
where each disjunct contain all possible values
for a given variable.
It is stored in a hash structure
such that for each variable
the hash function will return all possible
values for that variable.
This representation speeds up the processing then checking for
contradictory and tautological
local conditions.

Next, we present the operations of
the PossDB system.
Note that none of these operations affect the global condition.\\

{\bf Relational Selection}
The select statement generalizes the standard SQL
select statement. The generalized select statement
will work on c-tables
rather than relations with {\tt null}'s.
Beside returning the exact answer,
the select statement
also optimizes the c-table by removing tuples $t$,
where $\varphi(t)\wedge\Phi(T)$ is a contradiction,
and replacing  with {\em true} local conditions of tuples $t$,
where $\Phi(T)$ implies $\varphi(t)$.

Consider e.g.\ the query that returns all employee
from the 'IT' department:
\begin{verbatim}
SELECT * FROM Emp WHERE Dept = 'IT';
\end{verbatim}\\
The query results in the following
c-table:
\begin{center}
\begin{tabular}{llll|l}
Name     & Gender & Mstat   & Dept    & $\varphi(t)$  \\ \hline\hline 
Alice    & $x_1$  & married & IT      & {\em True}    \\
David    & M      & married & $x_4$   & $x_4=\mbox{'IT'}$   \\
Ella     & F       & single  & $x_4$   & $x_4=\mbox{'IT'}$
\end{tabular}
\end{center}

Note that the query returns a representation of the exact answer,
that is a c-table that represents the set of all possible answer instances.
This pertains to all query operations in the PossDB system.\\

{\bf Relational Projection}
operation is implemented, as expected,
as an extension
of the {\tt SELECT} statement.\\

{\bf Relational Join}
The join and cross product operations
work similarly with their standard SQL counterparts.
For example consider the following
project-join query
that returns all pairs of employee names
that work in the same department such that
the first employee is male and the second employee
is female:
\newline
\verb+SELECT e1.Name as Name1, e2.Name as Name2+
\newline
\verb+FROM Emp e1 INNER JOIN Emp e2 ON e1.Dept=e2.Dept+
\newline
\verb+WHERE e1.Gender='M' AND e2.Gender='F'+

\smallskip
The exact answer for this query is:
\begin{center}
\begin{tabular}{ll|l}
\multicolumn{3}{c}{Q} \\ \hline
Name1 & Name2 & $\varphi(t)$\\ \hline\hline
Alice & Ella & $x_1=\mbox{'M'} \wedge x_4=\mbox{'IT'}$\\
Bob & Cecilia & $x_2=\mbox{'M'} \wedge x_3=\mbox{'F'}$\\
\color{red}\sout{Bob} & \color{red}\sout{Ella} &
\color{red}\sout{$x_2=\mbox{'M'} \wedge x_4=\mbox{'HR'}$} \\
Cecilia & Bob & $x_3=\mbox{'M'} \wedge x_2=\mbox{'F'}$\\
\color{red}\sout{Cecilia} & \color{red}\sout{Ella} &
\color{red}\sout{$x_3=\mbox{'M'} \wedge x_4=\mbox{'HR'}$} \\
David & Alice & $x_1=\mbox{'F'} \wedge x_4=\mbox{'IT'}$ \\
\color{red}\sout{David} & \color{red}\sout{Bob} &
\color{red}\sout{$x_2=\mbox{'F'} \wedge x_4=\mbox{'HR'}$} \\
\color{red}\sout{David} & \color{red}\sout{Cecilia} &
\color{red}\sout{$x_3=\mbox{'F'} \wedge x_4=\mbox{'HR'}$} \\
David & Ella & {\em True}
\end{tabular}
\end{center}

\smallskip

It can be noted that in the join case the local conditions
for each resulted tuple is a conjunction of the local conditions of
the tuples that were joined, and the
condition induced by the ``where" clause in the select statement.
The overstriked tuples are deleted by the system,
since they have local conditions that are not
satisfiable together with the global condition.
Also note that the local condition for the last tuple is
a tautology as both employees ``David" and ``Ella"
share the same variable as department.\\

{\bf Insertion}
C-SQL extends the standard SQL Insert statement
by allowing the users to also specify a local
condition associated with the inserted tuple.
In case the {\tt CONDITION} clause is not specified in the
{\tt INSERT} statement,
by default we consider the local condition tautological, i.e.\ {\em True}.
The following example shows the syntax used to insert the tuple
(Smith, M, single, $x$) with the local condition
``$x = $HR or $x = $PR" in the
Emp table.\\

\texttt{INSERT INTO Emp}

\texttt{VALUES ('Smith','M','single',x) }

\texttt{CONDITION ($x = $'HR' or $x = $'PR')}

\bigskip

{\bf Query Answers}
We return the exact answer as a c-table.
This is comparable with MayBMS-1 that returns
all the tuples that can occur in the query answer
or some complete instance corresponding the input database.
This has the drawback that the answer may contain two mutually exclusive
tuples. On the other hand PossDB returns a c-table
representing the exact answer. In some cases this c-table
might have convoluted local conditions, and it might be
difficult for the user to understand the structure.
In order to overcome this, we have included two additional
functions {\tt IS POSSIBLE} and {\tt IS CERTAIN}. Both functions
work in polynomial time.\\

{\bf Special functions}
We have two new functions unique to PossDB.
These functions are used to query
for certainty and possibility of a tuple in a c-table or
in the result of a query.

\smallskip
\texttt{IS POSSIBLE(Tuple) IN C-Table Name | Query }
\smallskip

The {\tt IS POSSIBLE} is a Boolean function takes a tuple {\tt Tuple} and
decides if the tuple is possible in the c-table given
by the {\tt C-Table Name} or in the result of the given {\tt Query}.
Intuitively a tuple is possible in a given
c-table if there exists a valuation for the c-table
that contains that tuple.
 The ``Tuple" has to be specified as a list of
(Name, Value) pairs.
As an example consider the following function call:

\texttt{IS POSSIBLE(Name, 'Bob', Gender, 'M',}

\texttt{~~~~~~~Mstat, 'married', Dept, 'HR') IN Emp}

\smallskip
With the data from the previous example
the {\tt IS POSSIBLE} function returns {\em true},
because given tuple is possible in the system.
However, it is not certain
because it depends on the condition $(x_2 = \mbox{'M'})$.

\texttt{IS CERTAIN(Tuple) IN C-Table Name | Query}

\smallskip

Similarly to {\tt{IS POSSIBLE}} the {\tt{IS CERTAIN}} function takes as parameter a
tuple, and  a c-table name or query
and returns {\em True} if the tuple is certain in the
given c-table or the result of the given query.
Certain means that the tuple
appears under all possible interpretations of the nulls.
The following is an example of the
usage of the {\tt{IS CERTAIN}} function

\texttt{IS CERTAIN (Name, 'Bob', Dept, 'HR') IN}

\texttt{~SELECT Name, Dept FROM Emp}

\smallskip

This function returns {\em True},
because given tuple appear under any interpretation for
the nulls. Note also that the function would return
{\em False} if
we also included the Gender column in the query.

This could be extended so that the user could ask
if a {\em set} of tuples is possible or certain. Thus we
could also determine whether two possible tuples
are mutually exclusive, by issuing {\tt{IS POSSIBLE}} $t_1$,
{\tt{IS POSSIBLE}} $t_2$, and {\tt{IS POSSIBLE \{$t_1,t_2$\}}}. If
the first two answers are $True$ and the third answer is $False$, it means
that both $t_1$ and $t_2$ are possible tuples, but they are mutually exclusive (i.e.
cannot co-exist in the same possible world). We note that the {\tt{IS CERTAIN}}
would still run in polynomial time
in this generalization, as would also the {\tt{IS CERTAIN}} function, provided the number of tuples in the set were fixed \cite{DBLP:journals/tcs/AbiteboulKG91}.

\section{Implementation}
Without loss of generality, the information in our conditional
tables are encoded as integers. Positive integers
denote constants and negative integers denote nulls.
Consequently, without variables, the PossDB
system works as a regular RDBMS,
and the performance in this case will be the same as
that of PostgreSQL.
With this encoding we need to be sure that
the queries are properly evaluated.
Thus,
each equality condition of the form $A=c$
part of a C-SQL query, where $A$ is a column name and
$c$ a constant,
is rewritten as $(A=c\vee A<0)$ in SQL. This is necessary in
order to check that the column $A$ is either constant $c$
or that it represents
a null value, here encoded as negative integers.
In order to check for satisfiability of a local conditions and
its conjunction with the global condition,
the local conditions are converted into
DNF (Disjunctive Normal Form).
To make a faster satisfiability
test we store the global condition as hash based representation of its CNF (Conjunctive Normal Form).

After the Satisfiability and Tautology checks,
the system decides which tuple to show in the
result of the query by adding some annotations in the
local condition column.
Our application has a GUI
capable of generating the query result in a human readable interface.
From a technical perspective PossDB
is a two-layer system, the application layer built in Java
and as a database layer it uses PostgreSQL database engine.
When the user types a C-SQL query,
the system interprets it and execute it by a series of SQL statements
against the database and a series of Java calls needed to
make sure that the c-tables are correctly manipulated
and displayed to the user.

{\bf System Architecture}.
PossDB system is built on top of PostgreSQL.
On the middle tier Java\textregistered~ and ANTLR\cite{Parr95antlr:a}
are being used. ANTLR is used to parse the C-SQL queries and database conditions,
while Java is used to implement the C-SQL processing part,
displaying the results,
evaluating conditions, and connecting to the PostgreSQL database server.

This Java application is working
with input and output streams, hence it can be easily ported to the any kind
of application server or simply used through a console.
The connection between the Java middle tier and PostgreSQL database server is done through JDBC.

\begin{figure}[ht]
    \centering
    \includegraphics[width=0.47\textwidth,natwidth=300,natheight=300]{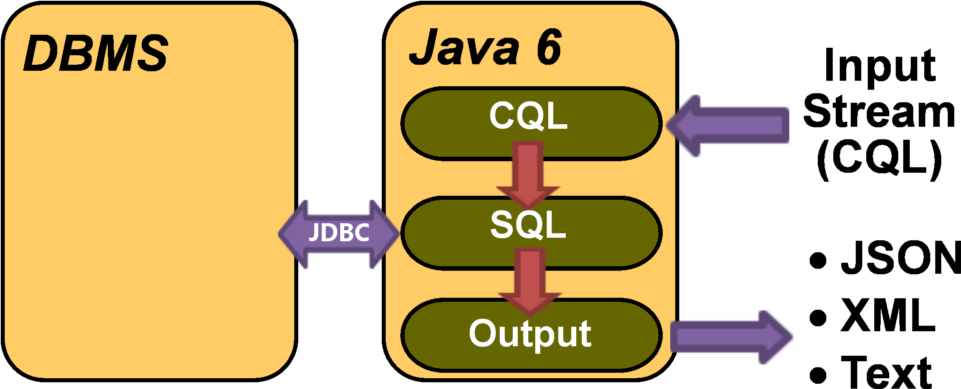}
    \caption[]{System Workflow}
\end{figure}

\section{Experimental Results}

In order to check the scalability of
our system we wanted to compare it with another system
that has the most similar features.
Thus we compared PossDB with MayBMS-1,
as MayBMS-1 also returns the exact answer to queries,
and the scalability of MayBMS has been proven
\cite{Antova:2009:WBE:1644245.1644253}.
Furthermore, both PossDB and MayBMS-1 are built
on top of PostgeSQL.

Our experiments are based on the queries and data
which were used for the MayBMS-1 experimental evaluation
\cite{Antova:2009:WBE:1644245.1644253}.
Their experiment used a large census database encrypted as integers
\cite{Ruggles2004}.
Noise was introduced by replacing some values with
variables that could take between 2 and 8 possible values.
A noise ratio of $n\%$ meant that $n\%$ of the values
were perturbed in this fashion.
In our experiments we used to same data
and noise generator as MayBMS-1.
The MayBMS-1 system and
the noise generator were obtained from \cite{Doe:2009:Misc}. 

We tested both systems with up to 10 million tuples. The charts below
contain the result of the test using queries $Q_1$ and $Q_2$
from the experiments in
\cite{Antova:2009:WBE:1644245.1644253}.
The results show that PossDB clearly outperforms MayBMS-1.

{\bf System configuration}.
We conducted all our experiments on
Intel\textregistered Core\texttrademark i5-760 processor
machine with 8 GB RAM, running Windows 7 Enterprise and PostgreSQL 9.0.

\begin{figure*}[htb]
  
   \includegraphics[width=0.3\textwidth]{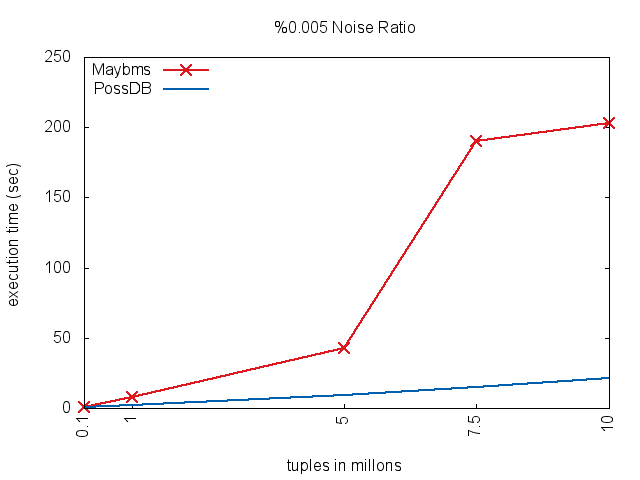}
   \includegraphics[width=0.3\textwidth]{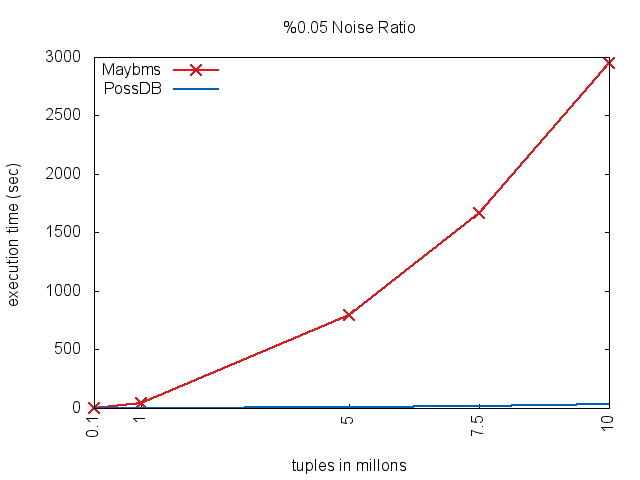}
   \includegraphics[width=0.3\textwidth]{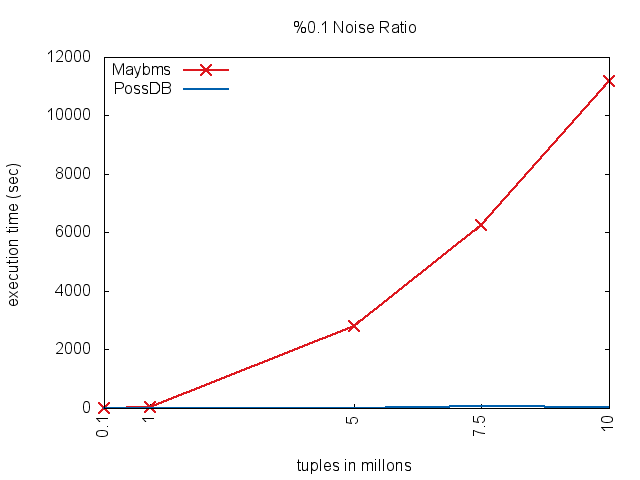}\\
    Q1: {\small{\tt SELECT * FROM R WHERE VETSTAT = 8 AND  CITIZEN = 9}}

   \includegraphics[width=0.3\textwidth]{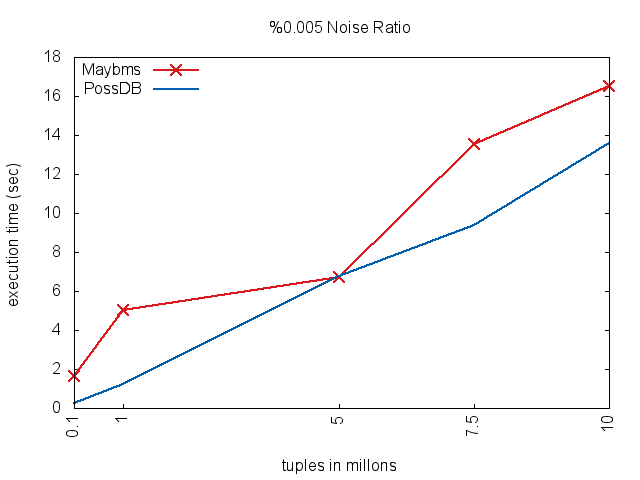}
   \includegraphics[width=0.3\textwidth]{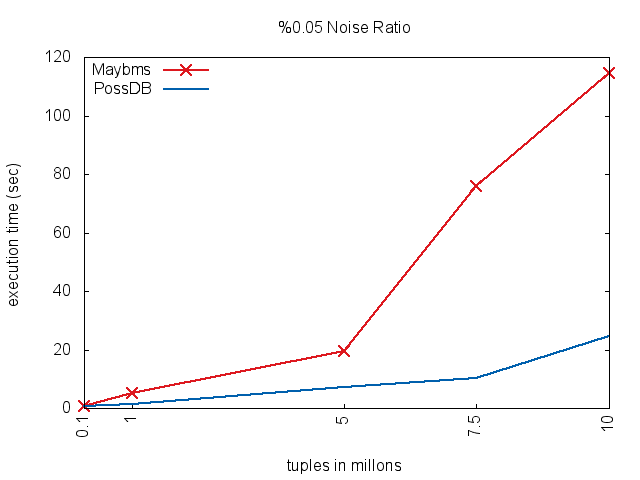}
   \includegraphics[width=0.3\textwidth]{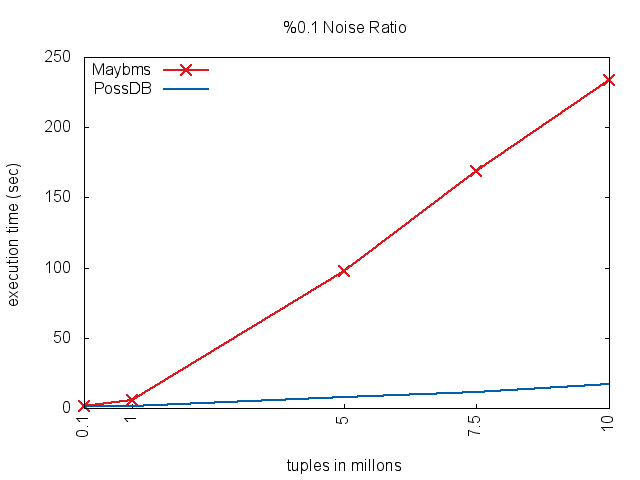}\\
   Q2: {\small{\tt SELECT STATEFIP,OCC1990,CITIZEN,SUBFAM FROM R }}

       {~~~~~~\small{\tt WHERE STATEFIP = OCC1990 AND CITIZEN = 1 AND SUBFAM > 4 }}

\end{figure*}

%

\section{Conclusions \& Further Work}
The system presented here is capable to store incomplete
data using Conditional Tables. This structure, even if well known,
has not been implemented before, although many probabilistic systems
essentially use probabilistic versions of c-tables.

In this report we show not only that the conditional table is a good
candidate
for storing incomplete information but we also show that that the
system indeed is scalable. For now PossDB is able to
process positive queries. We are in the process of extending the system
to allow general SQL queries, including also certain/possible nested
subqueries. This requires non-trivial extensions to the current C-SQL
language.
Another extension is to integrate a state-of-the-art SAT-solver,
e.g. \cite{DBLP:conf/ijcai/AudemardS09} or
\cite{DBLP:conf/hvc/Biere11}. The SAT-solver would then handle
the satisfiability and tautology tests, which is likely to
further improve the performance of the system.
Finally, we will also  implement the chase based procedure on
conditional tables \cite{DBLP:conf/lid/GrahneO11}
in order for the new system to be also usable in other applications,
such as Data Exchange, Data Repair and Data Integration.
%
%
\bibliographystyle{abbrv}
\bibliography{possdb}

\end{document}